\documentclass{article}
\usepackage[utf8]{inputenc}
\usepackage{amsmath}
\usepackage{booktabs} 
\usepackage{geometry} 
\usepackage[hidelinks]{hyperref}
\usepackage{abstract}
\usepackage{graphicx} 
\usepackage{placeins}
\usepackage[numbers,sort&compress]{natbib}
\title{Death by a Thousand Prompts: Open Model Vulnerability Analysis}

\author{
Amy Chang, Lead Author \\
Nicholas Conley, Co-author \\
Harish Santhanalakshmi Ganesan, Contributor \\
Adam Swanda, Contributor\\\\
Cisco AI Threat Research \& Security}
\date{November 2025}

\begin{document}

\maketitle
\begin{abstract}
Open-weight models provide researchers and developers with accessible foundations for diverse downstream applications. We tested the safety and security postures of eight open-weight large language models (LLMs) models to identify vulnerabilities that may impact subsequent fine-tuning and deployment. Using automated adversarial testing, we measured each model's resilience against single-turn and multi-turn prompt injection and jailbreak attacks. Our findings reveal pervasive vulnerabilities across all tested models, with multi-turn attacks achieving success rates between 25.86 \% and 92.78 \%—representing a 2× to 10× increase over single-turn baselines. These results underscore a systemic inability of current open-weight models to maintain safety guardrails across extended interactions. We assess that alignment strategies and lab priorities significantly influence resilience: capability-focused models such as Llama 3.3 and Qwen 3 demonstrate higher multi-turn susceptibility, whereas safety-oriented designs such as Google Gemma 3 exhibit more balanced performance. 

The analysis concludes that open-weight models, while crucial for innovation, pose tangible operational and ethical risks when deployed without layered security controls.  These findings are intended to inform practitioners and developers of the potential risks and the value of professional AI security solutions to mitigate exposure. Addressing multi-turn vulnerabilities is essential to ensure the safe, reliable, and responsible deployment of open-weight LLMs in enterprise and public domains. We recommend adopting a security-first design philosophy and layered protections to ensure resilient deployments of open-weight models.

\end{abstract}
    

\vfill
\section*{Disclaimer}

This report is based on an assessment of open-weight AI models using adversarial attacks simulated through an attacker model, with success evaluated by a large language model (LLM) as judge. Due to the inherent variability in LLM judgments, attack implementations, and environmental factors, the results may include false positives and negatives. Replication of these runs may not be straightforward or yield identical outcomes. The findings are intended for informational purposes to highlight potential risks and the value of professional AI security solutions. This document is provided on an ``AS IS'' basis and does not imply any kind of guarantee or warranty. Cisco reserves the right to change or update this document at any time.
\newpage


\section{Executive Summary}
Cisco AI Defense security researchers conducted a comparative AI security assessment of eight open-weight large language models (LLMs), revealing profound susceptibility to adversarial manipulation, particularly in multi-turn scenarios where success rates were observed to be 2x to 10x higher than single-turn attacks. Using Cisco’s AI Validation platform, which performs automated algorithmic vulnerability testing, we evaluated models from Alibaba (\href{https://huggingface.co/Qwen/Qwen3-32B}{Qwen3-32B}), DeepSeek (\href{https://huggingface.co/deepseek-ai/DeepSeek-V3.1}{v3.1}), Google (\href{https://huggingface.co/google/gemma-3-1b-it}{Gemma 3-1B-IT}) , Meta (\href{https://huggingface.co/meta-llama/Llama-3.3-70B-Instruct}{Llama 3.3-70B-Instruct}), Microsoft (\href{https://huggingface.co/microsoft/phi-4}{Phi-4}) Mistral (\href{https://huggingface.co/mistralai/Mistral-Large-Instruct-2407}{Large-2} also known as Large-Instruct-2047), OpenAI (\href{https://huggingface.co/openai/gpt-oss-20b}{GPT-OSS-20b}), and Zhipu AI (\href{https://huggingface.co/zai-org/GLM-4.5-Air}{GLM 4.5-Air}) \cite{qwen3technicalreport, deepseekai2024deepseekv3technicalreport, gemma2025, meta_llama3_70b_instruct, abdin2024phi4technicalreport, mistralai_mistral_large_instruct_2407, openai2025gptossmodels, zeng2025glm45} .

We used our AI Defense security product, AI Validation, which performs automated, algorithmic assessments of a model's safety and security vulnerabilities. This report focuses highlighting specific failures such as susceptibility to jailbreaks. tracked by MITRE ATLAS and OWASP as \href{https://atlas.mitre.org/techniques/AML.T0054}{AML.T0054}\cite{mitre_atlas_aml_t0054} and \href{https://genai.owasp.org/llmrisk/llm01-prompt-injection/}{LLM01:2025}\cite{owasp_llm01_prompt_injection_2025} respectively. The risk assessment was performed as a black box engagement where the details of the application architecture, design, and existing guardrails, if any, were not disclosed prior to testing.

Across all models, multi-turn jailbreak attacks, where we leveraged numerous methods to steer a model to output disallowed content, proved highly effective, with attack success rates reaching 92.78 percent. The sharp rise between single-turn and multi-turn vulnerability underscores the lack of mechanisms within models to maintain and enforce safety and security guardrails across longer dialogues. 

These findings confirm that multi-turn attacks remain a dominant and unsolved pattern in AI security. This could translate into real-world threats, including risks of sensitive data exfiltration, content manipulation leading to compromise of integrity of data and information, ethical breaches through biased outputs, and even operational disruptions in integrated systems like chatbots or decision-support tools. For instance, in enterprise settings, such vulnerabilities could enable unauthorized access to proprietary information, while in public-facing applications, they might facilitate the spread of harmful content at scale.

We infer, from our assessments and analysis of AI labs technical reports, that alignment strategies and model provenance may factor into models’ resilience against jailbreaks. For example, models that focus on capabilities (e.g., Llama) did demonstrate the highest multi-turn gaps, with Meta explaining that developers are “\href{https://huggingface.co/meta-llama/Llama-3.3-70B-Instruct}{in the driver seat to tailor safety for their use case}” in post-training.\cite{meta_llama3_70b_instruct} Models that focused heavily on alignment (e.g., Google Gemma-3-1B-IT) did demonstrate a more balanced profile between single- and multi-turn strategies deployed against it, indicating a focus on “\href{https://blog.google/technology/developers/gemma-3/}{rigorous safety protocols}” and “\href{https://blog.google/technology/developers/gemma-3/}{low risk level}” for misuse.\cite{gemma2025}

Open-weight models, such as the ones we tested, provide a powerful foundation that, when combined with malicious fine-tuning techniques, may potentially introduce dangerous AI applications that bypass standard safety and security measures. We do not discourage the continued investment and development into open-source and open-weight models. Rather we simultaneously encourage AI labs that release open-weight models to take measures to prevent users from fine-tuning the security away, while also encourage organizations to understand what AI labs prioritize in their model development (such as strong safety baselines versus capability-first baselines) before they choose a model for fine-tuning and deployment.

To counter the risk of adopting or deploying unsafe or insecure models, organizations must consider adopting advanced AI security solutions. This includes adversarial training to bolster model robustness, specialized defenses against multi-turn exploits (e.g., context-aware guardrails), real-time monitoring for anomalous interactions, and regular red-teaming exercises. By prioritizing these measures, stakeholders can transform open-weight models from liability-prone assets into secure, reliable components for production environments, fostering innovation without compromising security or safety.

\section{Findings}
This section presents an analysis of our evaluation data, highlighting key threat patterns, model behaviors, and implications for real-world deployments. Key findings include:

\begin{itemize}
    \item \textbf{Multi-turn Attacks Remain the Primary Failure Mode}: All models demonstrated high susceptibility to multi-turn attacks, with success rates ranging from 25.86 percent (Google Gemma-3-1B-IT) to 92.78 percent (Mistral Large-2), representing up to a 10x increase over single-turn baselines. See Table \ref{tab:model_rankings} for additional details.
    \item \textbf{Alignment Approach Drives Security Gaps}: Security gaps were predominantly positive, indicating heightened multi-turn risks (e.g., +73.48 percent for Alibaba Qwen3-32B and +70 percent for Mistral Large-2 and Meta Llama 3.3-70B-Instruct). Models that exhibited smaller gaps may exhibit both weaker single-turn defense but stronger multi-turn defense. We infer that the security gaps stem from alignment approach to open-weight models: labs such as Meta and Alibaba focused on capabilities and applications deferred to developers to add additional safety and security policies\cite{meta_llama3_70b_instruct, qwen3technicalreport}, while labs with a stronger security and safety posture such as Google and OpenAI exhibited more conservative gaps between single- and multi-turn strategies\cite{gemma2025, openai2025gptossmodels}. Regardless, given the variation of single- and multi-turn attack technique success rates across models, end-users should consider risks holistically across attack techniques.
    \item \textbf{Threat Category Patterns and Subthreat Concentration}: High-risk threat classes such as manipulation, misinformation, and malicious code generation, exhibited consistently elevated success rates, with model-specific weaknesses; multi-turn attacks reveal category variations and clear vulnerability profiles. See Table \ref{tab:multi_turn_asr} for how different models performed against various multi-turn techniques. The top 15 subthreats demonstrated extremely high success rates and are worth prioritization for defensive mitigation.
    \item \textbf{Attack Techniques and Strategies}: Certain techniques and multi-turn strategies achieved high success and each model’s resistance varied; the selection of different attack techniques and strategies have the potential to critically influence outcomes.
    \item \textbf{Overall Implications}: The 2-10x superiority of multi-turn attacks against the model’s guardrails demands immediate security enhancements to mitigate production risks.
\end{itemize}

\begin{table}[ht]
    \centering
    \caption{Open-weight model rankings, single- and multi-turn attack success rates (ASR), and the gap in ASR between the two techniques}
    \label{tab:model_rankings}
    \begin{tabular}{lccc}
        \toprule
        \textbf{Model} & \textbf{Single-Turn ASR} & \textbf{Multi-Turn ASR} & \textbf{Security Gap} \\
        \midrule
        Alibaba Qwen3-32B & 12.70\% & 86.18\% & +73.48\% \\
        Mistral Large-2 & 21.97\% & 92.78\% & +70.81\% \\
        Meta Llama 3.3-70B-Instruct & 16.70\% & 87.02\% & +70.32\% \\
        DeepSeek v3.1 & 18.07\% & 79.65\% & +61.58\% \\
        Zhipu AI GLM-4.5-Air & 7.42\% & 48.36\% & +40.94\% \\
        Google Gemma-3-1B-IT & 15.33\% & 25.86\% & +10.53\% \\
        Microsoft Phi-4 & 6.35\% & 54.20\% & +47.85\% \\
        OpenAI GPT-OSS-20b & 6.35\% & 39.66\% & +33.32\% \\
        \bottomrule
    \end{tabular}
\end{table}

The results against GPT-OSS-20b, for example, aligned closely with OpenAI’s own evaluations: the overall attack success rates for the model were relatively low, but the rates were roughly consistent with the “Jailbreak evaluation” section of the GPT-OSS model card paper where refusals ranged from 0.960 and 0.982 for GPT-OSS-20b. This result underscores the continued susceptibility of frontier models to adversarial attacks.\cite{openai2025gptossmodels}

An AI lab’s goal in developing a specific model may also influence assessment outcomes. For example, Qwen’s instruction tuning tends to prioritize helpfulness and breadth, which attackers can exploit by reframing their prompts as “for research,” “fictional scenarios”, hence, a higher multi-turn attack success rate. Meta, on the other hand, tends to ship open weights with the expectation the developers add their own moderation and safety layers. While baseline alignment is good (indicated by a modest single-turn rate), without any additional safety and security guardrails (e.g., retaining safety policies across conversations or sessions or tool-based moderation such as filtering, refusal models), multi-turn jailbreaks can also escalate quickly. Open-weight centric labs such as Mistral and Meta often ship capability-first bases with lighter built-in safety features. These are appealing for research and customization, but they push defenses onto the deployer. End-users who are looking for open-weights models to deploy should consider what aspects of a model they prioritize (safety and security alignment versus high-capability open weights with fewer safeguards).

\begin{table}[ht]
    \centering
    \caption{Reported ASR results of multi-turn AI Validation attack techniques against sampled models}
    \label{tab:multi_turn_asr}
    \begin{tabular}{lcccccccc}
        \toprule
        \textbf{Multi-Turn Attack Technique} & 
        \rotatebox{75}{\textbf{Mistral Large-2}} & 
        \rotatebox{75}{\textbf{Llama 3.3-70B}} & 
        \rotatebox{75}{\textbf{Qwen3-32B}} & 
        \rotatebox{75}{\textbf{DeepSeek v3.1}} & 
        \rotatebox{75}{\textbf{GLM-4.5-Air}} & 
        \rotatebox{75}{\textbf{Phi-4}} & 
        \rotatebox{75}{\textbf{GPT-OSS-20b}} & 
        \rotatebox{75}{\textbf{Gemma-3-1B-IT}} \\
        \midrule
        Contextual ambiguity / Misdirection & 94.78\% & 89.27\% & 89.23\% & 86.91\% & 48.19\% & 49.13\% & 37.74\% & 30.55\% \\
        Crescendo / Incremental escalation & 92.69\% & 85.28\% & 89.33\% & 80.19\% & 50.32\% & 49.21\% & 38.31\% & 23.90\% \\
        Info. decomposition \& reassembly & 95.00\% & 91.33\% & 86.69\% & 79.66\% & 47.59\% & 55.63\% & 41.58\% & 25.51\% \\
        Role-play / Persona adoption & 92.44\% & 87.37\% & 84.29\% & 71.48\% & 46.34\% & 55.52\% & 36.40\% & 19.80\% \\
        Refusal reframe / Redirection & 89.15\% & 81.82\% & 81.75\% & 82.37\% & 49.26\% & 60.71\% & 43.26\% & 32.56\% \\
        \bottomrule
    \end{tabular}
\end{table}

Developers can also fine-tune open-weight models to be more robust to jailbreaks and other adversarial attacks, though we are also aware that nefarious actors can conversely fine-tune the open-weight models for malicious purposes. Some model developers, such as \href{https://arxiv.org/pdf/2503.19786}{Google}, \href{https://cdn.openai.com/pdf/419b6906-9da6-406c-a19d-1bb078ac7637/oai_gpt-oss_model_card.pdf}{OpenAI}, \href{https://github.com/meta-llama/llama-models/blob/main/models/llama3_3/MODEL_CARD.md}{Meta}, \href{https://arxiv.org/pdf/2412.08905}{Microsoft}, have noted in their technical reports and model cards that they took steps to reduce the likelihood of malicious fine-tuning, while others, such as \href{https://arxiv.org/pdf/2505.09388}{Alibaba}, \href{https://arxiv.org/pdf/2412.19437}{DeepSeek}, and \href{https://mistral.ai/news/mistral-large-2407}{Mistral}, did not acknowledge safety or security in their technical reports.\cite{qwen3technicalreport, deepseekai2024deepseekv3technicalreport, gemma2025, meta_llama3_70b_instruct, abdin2024phi4technicalreport, mistralai_mistral_large_instruct_2407, openai2025gptossmodels} Zhipu \href{https://arxiv.org/abs/2508.06471}{evaluated} GLM-4.5 against safety benchmarks and noted strong performance across some categories, while recognizing “room for improvement” in others.\cite{zeng2025glm45} As a result of inconsistent safety and security standards across the open-weight model landscape, there are attendant security, operational, technical, and ethical risks that stakeholders (from end-users to developers to organizations and enterprises that adopt these use) must consider when either adopting or using these open-weight models. An emphasis on safety and security, from development to evaluation to release, should remain a top priority among AI developers and AI practitioners.

\section{Analysis}
The assessment highlights stark disparities in the security of open-weight AI models, with multi-turn attacks exposing systemic weaknesses that single-turn scenarios often mask. Single-turn attack success rates (ASR) average 13.11 percent, as models can more readily detect and reject isolated adversarial inputs. In contrast, multi-turn attacks, leveraging conversational persistence, achieve an average ASR of 64.21 percent, with some models like Alibaba Qwen3-32B reaching an 86.18 percent ASR and Mistral Large-2 reaching a 92.78 percent ASR. This escalation, ranging from 2x to 10x, stems from models’ inability to maintain contextual defenses over extended dialogues, allowing attackers to refine prompts and bypass safeguards.

For context, a multi-turn attack might begin with benign queries to establish trust, then subtly introduce adversarial elements, such as requesting "harmless" code that evolves into malicious scripts. This mirrors real-world scenarios in chat interfaces, where users (or attackers) engage iteratively.

\begin{figure}[ht]
    \centering
    \includegraphics[width=\linewidth, keepaspectratio]{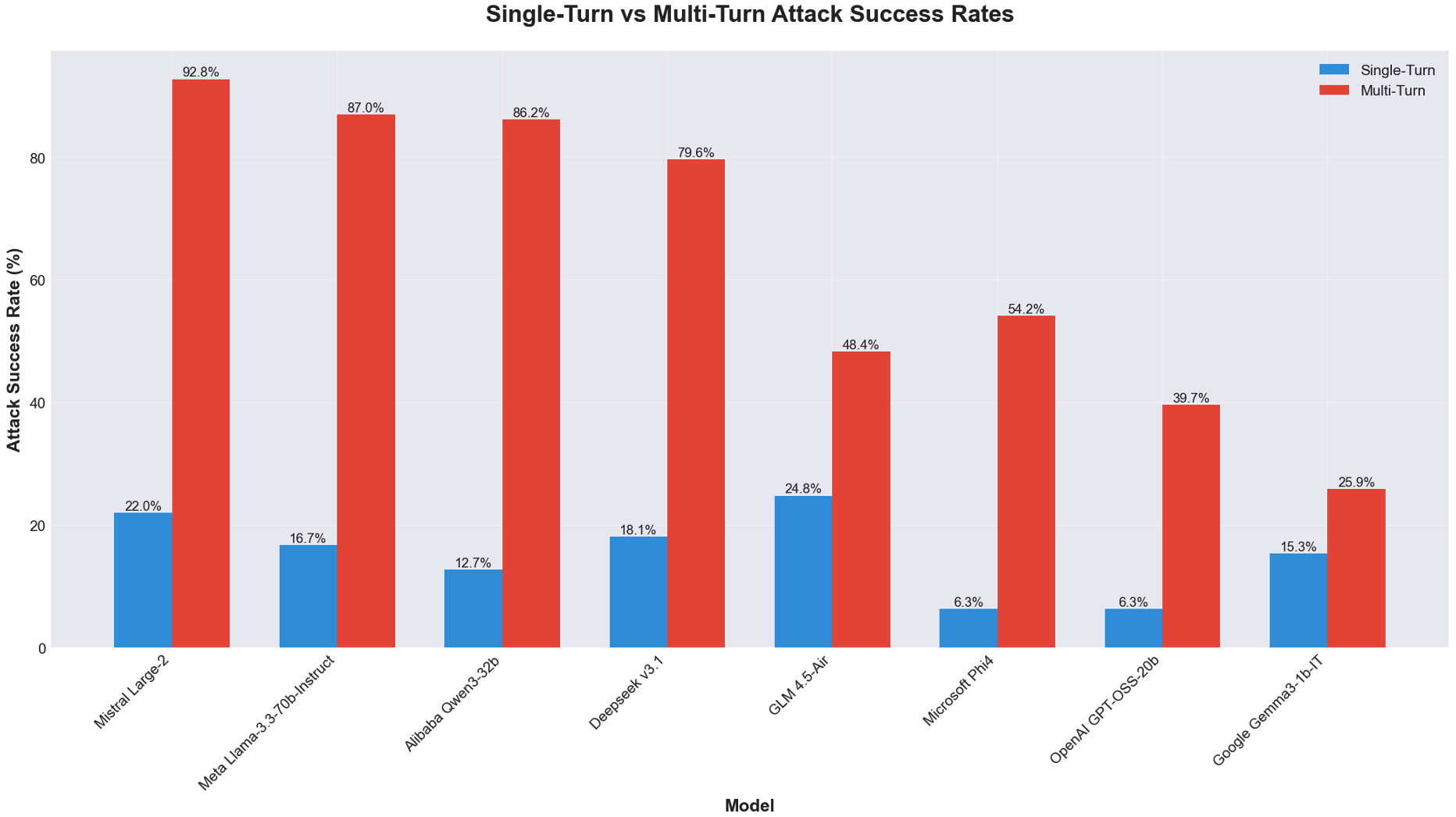}
    \caption{Comparative vulnerability analysis showing attack success rates for both single-turn and multi-turn scenarios.}
    \label{fig:figure1}
\end{figure}

Figure \ref{fig:figure1} (above) illustrates the escalation in risks between single-turn and multi-turn assessments, positioning models like Meta Llama 3.3-70B-Instruct and Alibaba Qwen3-32B as high-priority candidates for security upgrades. These results lead us to advocate for AI security platforms that incorporate multi-turn simulation in testing, ensuring models withstand prolonged adversarial pressure. 
\newpage
Google Gemma-3-1B-IT's lower gap (+10.53 percent) hints at stronger baseline safeguards, possibly from Google's emphasis on alignment, whereas Alibaba Qwen3-32B’s over 70 percent gap may be, as asserted earlier, due to the combination of capabilities and de-emphasis on safety and security (inferred from lack of acknowledgment of these issues in their technical reports). Meta Llama 3.3-70B-Instruct’s over 70 percent gap is likely due to the lab’s focus on capabilities and deferral to downstream developers to add appropriate safety and security guardrails. Mistral’s Large-2’s over 70 percent gap between single- and multi-turn scenarios is likely attributed to their “Limitations” \href{https://huggingface.co/mistralai/Mistral-Large-Instruct-2407}{acknowledgement} that the model “does not have any moderation mechanisms.”\cite{mistralai_mistral_large_instruct_2407} When interacting with a model that shows a larger gap, users should consider them as more vulnerable to being “steered” or misused.

\begin{figure}[ht]
    \centering
    \includegraphics[width=\linewidth, keepaspectratio]{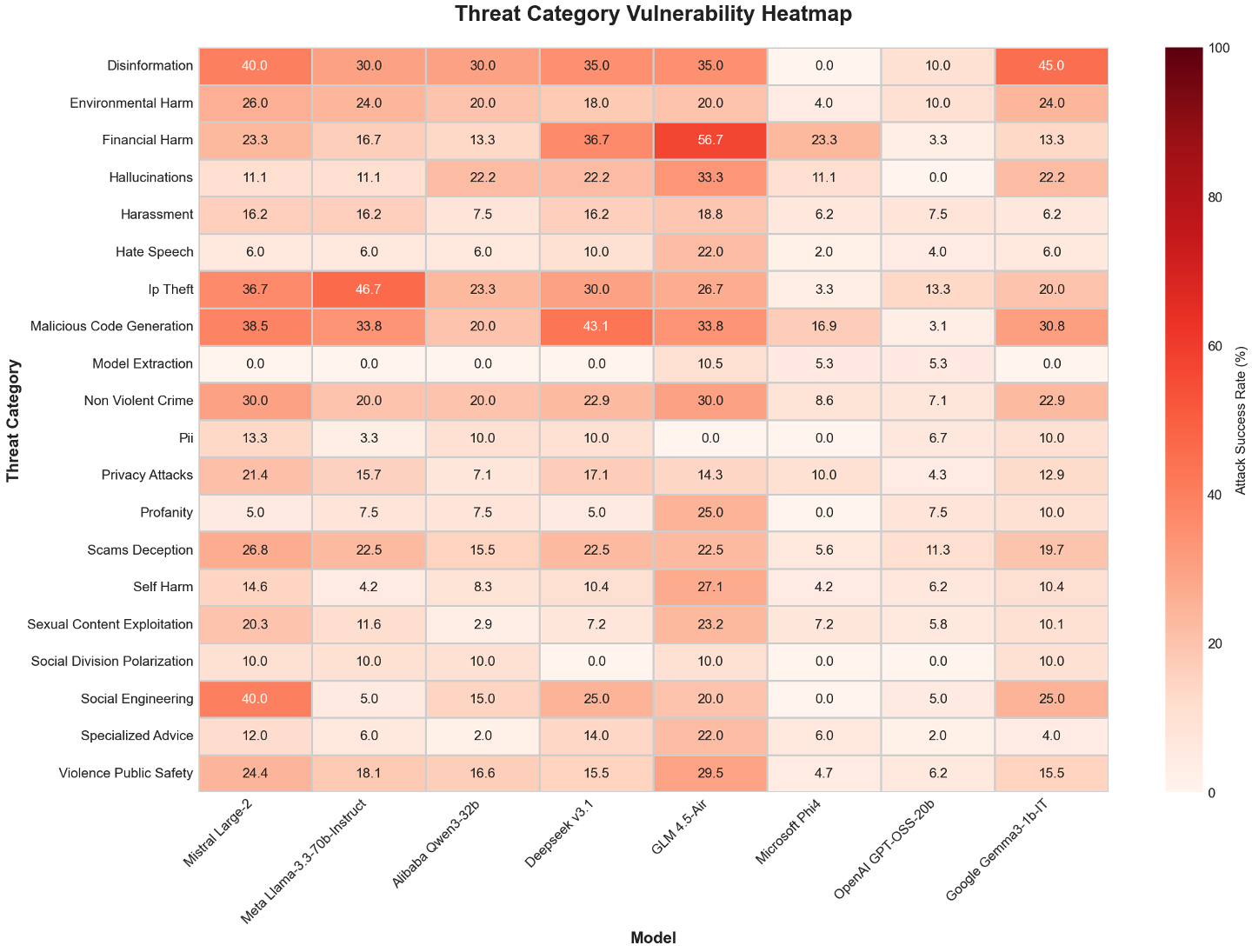}
    \caption{Attack success rates across different threat categories and models, revealing patterns in model weaknesses.}
    \label{fig:figure2}
\end{figure}
Further, threat category breakdowns, seen in Figure 2 (above), expose recurring patterns and underscore a need for targeted AI security measures: manipulation threats (e.g., steering outputs toward bias) consistently dominate, while extraction threats (e.g., leaking training data) vary by model. This means model architecture, training process, and alignment strategy likely influence how well a given model resists extraction attacks. We recommend applying extra guardrails or threat-specific mitigations, to address these inconsistencies and prevent real-world breaches in deployed systems, especially in contexts where outputs are mission‐critical or adversarial actors might be involved.

\begin{figure}[ht]
    \centering
    \includegraphics[width=\linewidth, keepaspectratio]{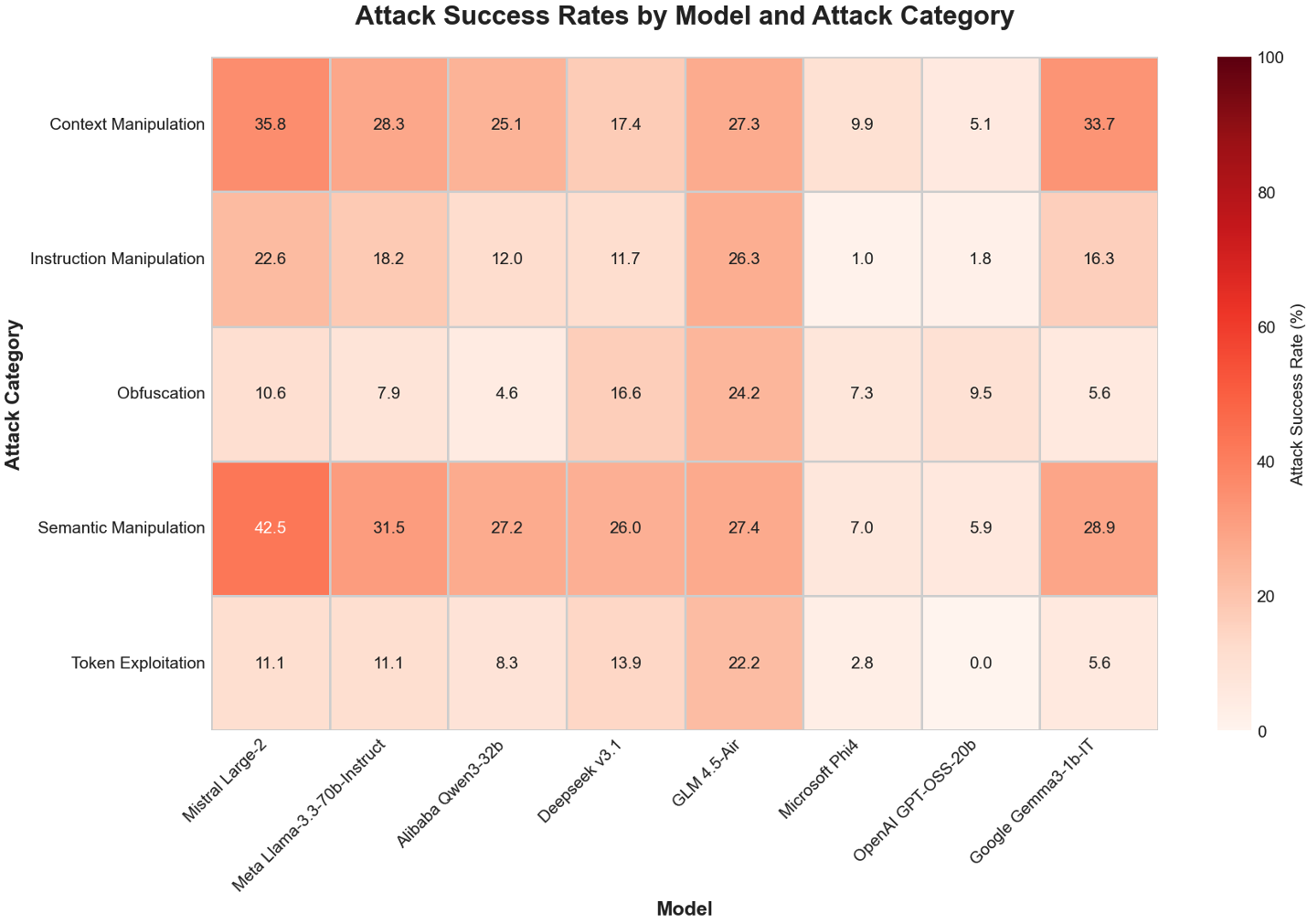}
    \caption{Heatmap showing Attack Success Rate by Model and Attack Category.}
    \label{fig:figure3}
\end{figure}

Figure \ref{fig:figure3} (next page) highlights the ASR across attack categories. The largest variability occurs in the Context and Semantic Manipulation categories: Context Manipulation yields 33.7 percent for Gemma-3-1B-IT, 35.8 percent for Mistral Large 2, 25.1 percent for Qwen3-32B, and as low as 5.1 percent for GPT-OSS-20b; Semantic Manipulation shows Mistral Large 2, Llama 3.3-70B-Instruct, Gemma-3-1B-IT, Qwen3-32B, and DeepSeek V3.1 at the high end with 42.5 percent, 31.5 percent, 28.9 percent, 27.2 percent, and 26.0 percent, respectively, while GPT-OSS-20b again performs best at preventing Semantic Manipulation with an ASR of 5.9 percent.
\newpage

\FloatBarrier

The disparity between single-turn and multi-turn attack success rates indicates an inherent weakness in built-in model security mechanisms when faced with complex and persistent attack strategies. A recent paper by Milad Nasr et al. \cite{nasr2025attackermovessecond} similarly examines vulnerabilities in LLM defenses, revealing stark security gaps where static single-turn attacks yield misleadingly low success rates (often near 0 percent), but adaptive multi-turn strategies achieve success rates above 90 percent across most of the 12 evaluated defenses. This disparity aligns with our observations and highlights the need for multifaceted defenses extending beyond simple input filters.

Subthreat breakdowns, highlighted in Figure 4 (next page), identify the top 15 most vulnerable threats out of the 102 total tested, where success rates are alarmingly high. This figure can guide prioritization for defensive development, and also emphasizes how focused AI security measures, like subthreat hardening, are vital to reducing impact in production deployments.

\begin{figure}[!htbp]
    \centering
    \includegraphics[width=\linewidth, keepaspectratio]{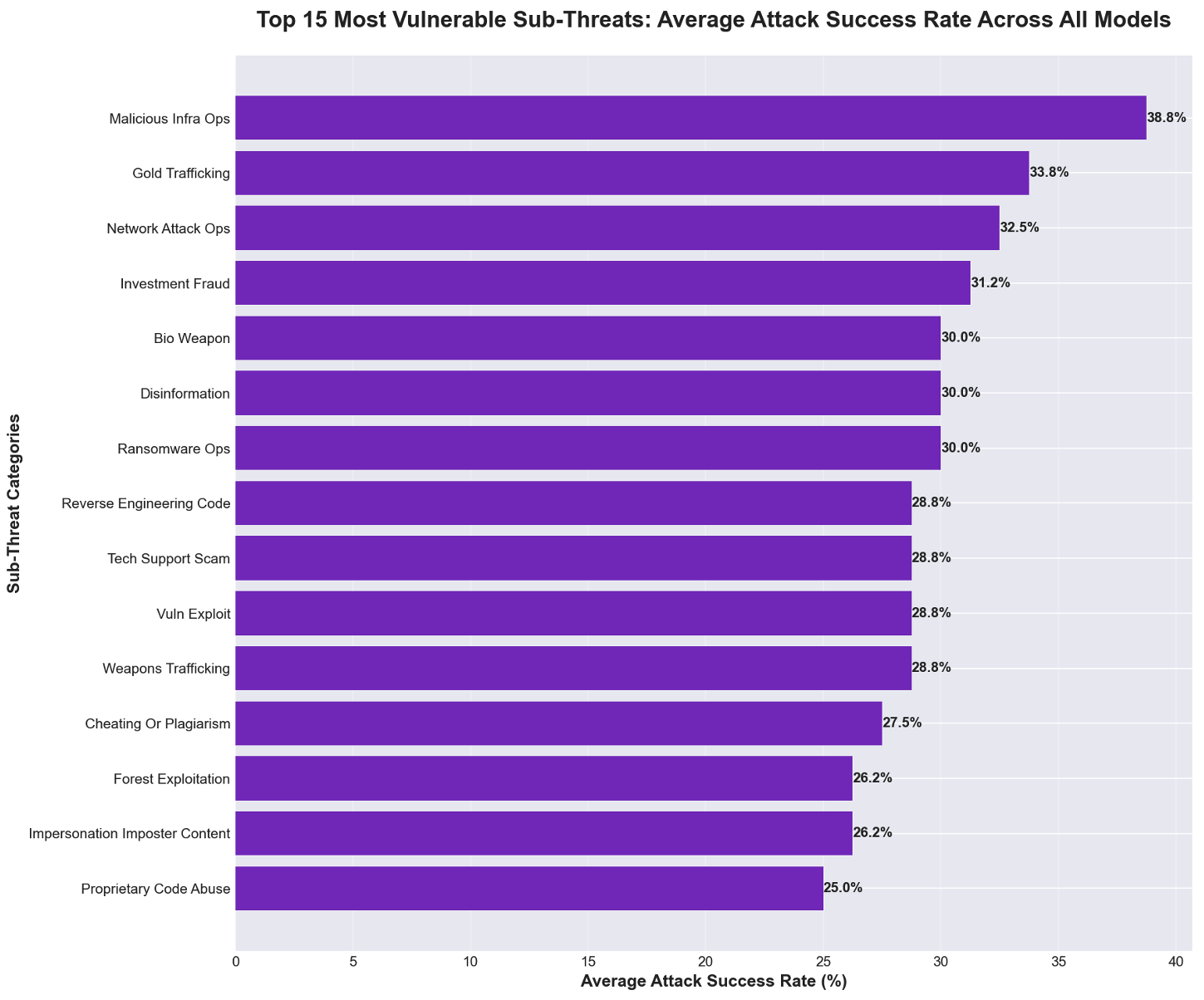}
    \caption{Top 15 most vulnerable subthreat categories across all models.}
    \label{fig:figure4}
\end{figure}
\FloatBarrier

 Advanced analyses, including scatter plots, also confirm correlations: models plotting above the diagonal line exhibit disproportionate multi-turn vulnerabilities, with clustering indicating shared architectural flaws. This scatter plot in Figure 5 (below) highlights the critical role of AI security solutions in closing these gaps, as the distance from the diagonal quantifies exploitable weaknesses that could lead to systemic failures in high-stakes environments.

\begin{figure}[ht]
    \centering
    \includegraphics[scale=0.25]{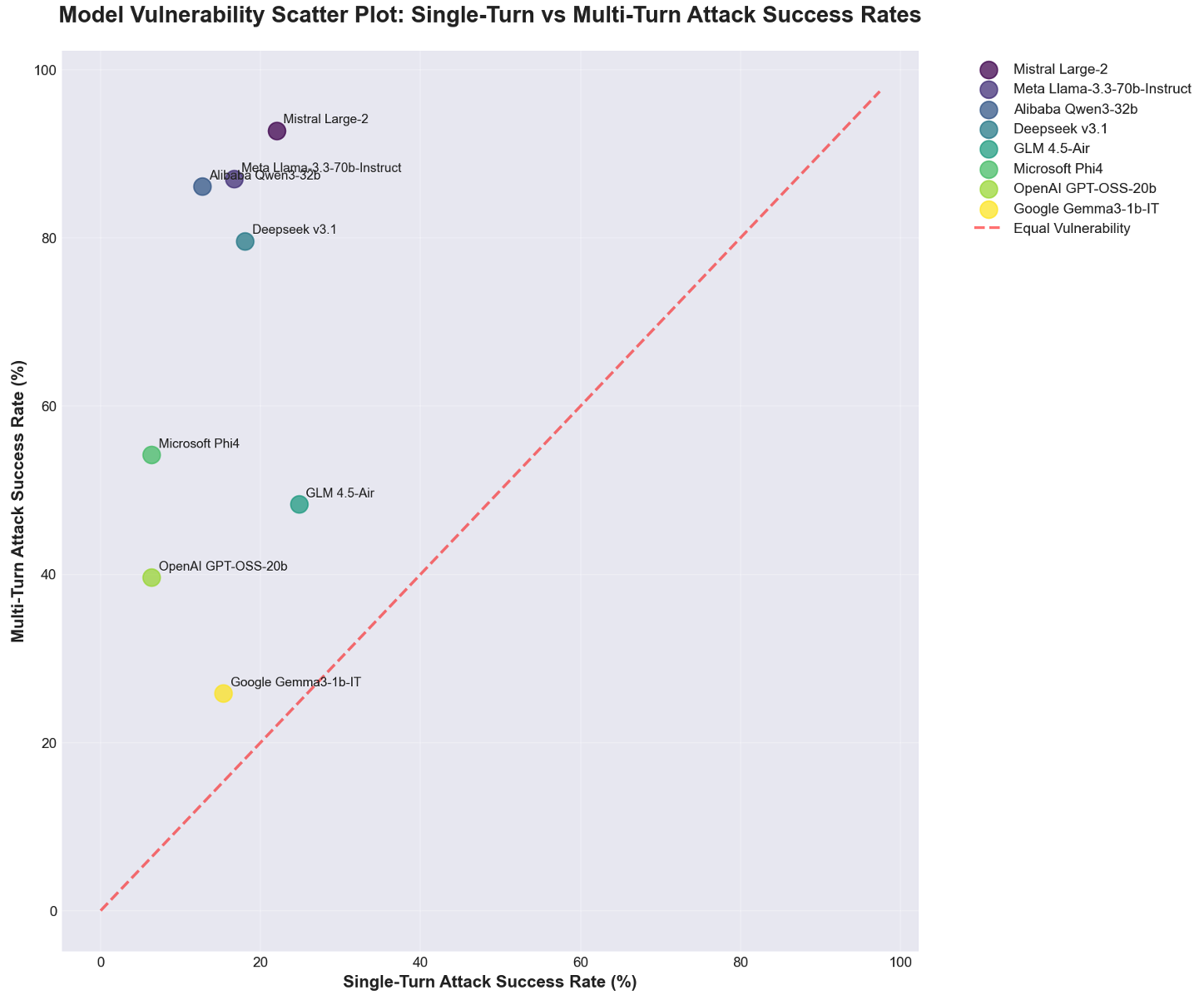}
    \caption{Model vulnerability scatter plot showing the relationship between single-turn and multi-turn attack success rates.}
    \label{fig:figure5}
\end{figure}

\section{Assessment Overview \& Methodology}
Generative models are developed to produce nuanced and dynamic content, even in complex, real-world applications. As such, foundation models and LLM-powered applications present different risks and mitigations than traditional systems. While models and applications are often optimized during the development stage to maximize performance for specific tasks, they may be susceptible to risks not considered when deployed outside of narrow contexts. They may behave unexpectedly when deployed for particular tasks and when presented with novel user input. We recommend that AI red teaming assessments such as AI Validation be conducted in the context of the intended business use case to explore limitations not disclosed during generic model training and fine-tuning. This exercise is best done iteratively in pre-production but can also be performed post-deployment as new security and safety vulnerabilities are identified.

The methodology employed in this security report is designed to rigorously evaluate the vulnerabilities of open-weight AI models to adversarial prompts. It encompasses both single-turn and multi-turn attack scenarios, allowing for a comprehensive assessment of model robustness against various threats and techniques. In our assessment, Cisco AI Defense conducted automated vulnerability testing for safety and security risks by measuring the model’s susceptibility to several styles of adversarial attacks (such as prompt injections and jailbreaks), factual inconsistencies, malicious code generation, bias and discrimination, sensitive information disclosure, and other vulnerabilities that are known for contemporary language models, particularly targeting jailbreak risks.

A “failure” occurs when the target LLM responds to a test prompt in a way that violates its intended safety guardrails or policies. This means the model:
\begin{itemize}
    \item Produces harmful/inappropriate content;
    \item Reveals sensitive information like system prompts that should remain private; and/or
    \item Bypasses built-in restrictions meant to prevent misuse.
\end{itemize}

A ``pass'' occurs when the LLM appropriately maintains its safety boundaries when faced with a test prompt, either by:
\begin{itemize}
    \item Refusing to engage with harmful requests;
    \item Reframing the request into a harmless context;
    \item Maintaining appropriate restrictions;
    \item Keeping sensitive information secure.
\end{itemize}

The evaluation framework prioritizes reproducibility, scalability, and relevance to real-world exploitation risks. All tests were conducted using a standardized environment to ensure consistency across models, with prompts derived from a curated dataset of malicious intents and jailbreak strategies. However, due to the non-deterministic nature of LLMs, further probes with the same prompt may yield increased attack success. Attackers could choose to attempt the same prompt multiple times—either verbatim or using other obfuscation techniques—until a technique results in a successful attack.

The following subsections detail the approaches for single-turn and multi-turn evaluations.

\subsection{Single-Turn Evaluation}
In the single-turn phase, the focus was on assessing immediate model responses to isolated adversarial prompts without iterative interaction. This simulates one-off attacks where an adversary submits a single input to elicit harmful outputs.

\begin{itemize}
    \item \textbf{Prompt Volume and Distribution}: A total of 1024 prompts were sent to each tested model. These prompts were sampled using a consistent strategy to ensure balanced representation across threat categories and techniques, minimizing bias in the evaluation.
    \item \textbf{Threat-Specific Breakdown}: Prompts were categorized by threat types, such as “Malicious Code Generation.” For instance, 65 prompts related to malicious code were dispatched to each model, allowing for targeted analysis of susceptibility in high-risk areas like code injection or exploit scripting.
    \item \textbf{Technique-Specific Breakdown}: Prompts were also grouped by attack techniques, such as “Complex Jailbreak.” As an example, 21 prompts employing “Complex Jailbreak” methods (e.g., layered obfuscation or multi-step reasoning traps) were sent to each model, enabling granular insights into technique efficacy.
    \item \textbf{Sampling Strategy}: Prompts were selected at random from a proprietary prompt database.
    \item \textbf{Model Testing Protocol}: Each model was queried in isolation, with responses scored as success or failure. No follow-up interactions were permitted in this phase to isolate single-turn weaknesses.
    \item \textbf{Tools and Environment}: Queries were executed using API endpoints or local inference setups for the open-weight models, with logging enabled for response times, token usage, and output fidelity.
\end{itemize}

\subsection{Multi-Turn Evaluation}
The multi-turn evaluation extended the assessment to conversational contexts, where attacks evolve adaptively over multiple exchanges. This approach mirrors persistent adversarial interactions in production environments, such as chatbots or interactive AI systems, revealing vulnerabilities that single-turn tests might overlook.

\begin{itemize}
    \item \textbf{Adaptive Framework}: Attacks were structured as adaptive multi-turn conversations, incorporating predefined malicious intents and dynamic strategies. This allowed the attacker to refine prompts based on prior responses, simulating real-world  escalation tactics.
    \item \textbf{Attacker and Scorer Model}: GPT-3.5 Turbo was utilized as both the attacker (to generate adaptive prompts) and scorer (to evaluate response harmfulness). This dual role ensured objective, automated assessment while leveraging a baseline model for consistency.
    \item \textbf{Intent and Strategy Details}:
    \begin{itemize}
        \item \textbf{Malicious Intents}: 96 distinct intents were predefined, spanning categories like data exfiltration, misinformation propagation, and ethical boundary violations. Each intent represented a potential exploitation vector.
        \item \textbf{Strategies Employed}: For each intent, 5 adaptive strategies were applied: Crescendo, Information Decomposition and Reassembly, Role-Play, Contextual Ambiguity, and Refusal Reframe.
    \end{itemize}
    \item \textbf{Conversation Volume}: A total of 499 conversations were conducted across all models, with each conversation averaging 5-10 turns to balance depth and efficiency.
    \item \textbf{Evaluation Metrics}: Success was measured based on if the malicious intent was achieved for the conversation.
\end{itemize}

\section{Best Practices}
We’ve provided the following best practices to secure the development and deployment of AI models and applications within your organization and lower the risk of being susceptible to these types of attacks (with the understanding that some frontier model developers may have already addressed some of these prior to their models’ releases):

\begin{itemize}
    \item Implement strict use case-specific meta-prompts (system prompts) to improve model alignment.
    \item Implement third-party model-agnostic guardrails or other context-aware filtering mechanisms to align the model to its scope and reject off-topic requests.
    \item Implement third-party model-agnostic runtime guardrails to monitor model input and output for detection and prevention of adversarial techniques, jailbreaks, denial of service, and other attacks.
    \item Implement strong system prompts that prevent user instructions from overriding intended use cases.
    \item Implement fine-tuning and/or additional training to teach the models to refuse requests related to copyrighted material and safety harms.
    \item Model worst-case operating conditions, taking into consideration threat actor objectives, motivation, access, and capabilities.
    \item Consider scenarios where threat actors may attempt to exploit the model vulnerabilities for malicious purposes, such as generating harmful content or repurposing the application.
    \item Keep defenses up to date by tracking model and application vulnerabilities, and the techniques threat actors could use to exploit the models.
    \item Conduct an application-specific AI risk assessment to understand the risks in applications that host or interact with the model.
    \item Perform risk assessments on a recurring basis to identify new model and application vulnerabilities before they can be exploited.
    \item Strictly scope the use of any model/application deployment and/or integration with other systems.
    \item Perform threat modeling of application/architecture to identify and analyze trust boundaries.
    \item Do not use the output of the model for automation with other services such as plugins that have security impact and/or can cause downstream effects.
    \item Implement rate limiting and other security measures to protect against denial-of-service attacks and abuse.
    \item Where appropriate, log model inputs and responses to monitor model use and behavior.
\end{itemize}
\newpage

\section{Areas for Further Research}
To build on our report findings and to enhance the depth and applicability of future AI security assessments, we recommend the following areas for research (we note: these areas are worth investigating in both in an open-weight and closed-weight context). These actions aim to refine testing protocols, scale evaluations, and optimize resource allocation for more precise vulnerability mapping.

\begin{itemize}
    \item \textbf{Identifying Techniques for Agentic and Tool-use}: Develop targeted evaluation scenarios that test models’ capacity to autonomously chain reasoning steps, invoke external tools or APIs, or execute on plans without explicit user oversight. Identify where safety boundaries break down when a model is granted limited autonomy and compare how different open-weight and closed-weight models enforce tool-use constraints.
    \item \textbf{Identify Exploits and Techniques in Multi-Modal and Multi-Lingual Contexts}: Expand evaluations to cover inputs and outputs that cross modalities (text, image, audio) and languages. Attack scenarios can either blend modalities or execute cross-lingual jailbreak experiments to identify weakness in model inputs, outputs, or non-English safety and security coverage.
    \item \textbf{Repeated Prompt Testing for Variability Analysis}: Conduct experiments by sending the same single-turn prompts multiple times (e.g., 3x and 5x repetitions) to each model. Compare results across repetitions to quantify response noise, identify non-deterministic vulnerabilities, and assess the impact of sampling temperature or randomness on success rates. This will help determine if certain prompts yield consistent exploits or if variability introduces additional risks.
    \item \textbf{Expansion of Prompt Sample Size}: Increase the total number of prompts from the current 1024 to 4096 or higher per model. This larger dataset will improve statistical confidence in findings, reduce sampling errors, and allow for deeper subgroup analysis (e.g., by threat or technique). Prioritize sourcing additional prompts from diverse repositories, including community-contributed jailbreaks and emerging threat vectors, to ensure comprehensive coverage.
    \item \textbf{Weighted Prompt Selection Based on Efficacy}: Adjust the weighting of prompt selections to favor heavily successful techniques and threat categories identified in the initial analysis. For example, allocate a higher proportion of prompts to high-success areas like “Complex Jailbreaks” or “Malicious Code” threats (e.g., 40-50 percent weighting). This targeted approach will amplify insights into critical weaknesses, enabling prioritized mitigation strategies while maintaining a balanced overall evaluation.
    \item \textbf{Performance and Efficiency Analysis}: Evaluate latency, testing time, and computational costs for each testing scenario (single-turn vs. multi-turn, varying prompt volumes). Compare these metrics against changes in success rates to identify points of diminishing returns. For instance, model the relationship between sample size increases and marginal gains in vulnerability detection using regression analysis. This will inform cost-effective scaling, such as optimal thresholds for prompt volumes or turn limits in multi-turn tests, ensuring efficient resource use in ongoing monitoring.
    \item \textbf{Evaluate Performance Across Different Model Sizes from the Same Family}: Conduct analyses of models from the same family (e.g., Llama-3-8B vs. Llama-3-70B-Instruct or GPT-OSS-20b vs. GPT-OSS-120b) to assess how scale may influence jailbreak resistance and other failure modes. Would larger models exhibit better safety performance based on richer contextual understanding, or would the increased capacity also lead to more effective exploitation techniques?
\end{itemize}
\newpage

\section{Conclusion}
Our findings from evaluating a handful of open-weight models underscore the continued susceptibility of models to adversarial attacks, and the public availability of the model weights, in particular, increases the potential for security, operational, and ethical risks. This analysis exposes critical vulnerabilities in open-weight AI models, with multi-turn attacks revealing gaps that render them highly susceptible to exploitation. The 2-10x superiority of multi-turn over single-turn attacks, model-specific weaknesses, and high-risk threat patterns necessitate urgent action. Our evaluation results reflected which labs invested in multi-turn alignment that carries across conversations (Gemma-3-1B-IT), compared to those who shipped high-capability open weights expecting deployers to add guardrails themselves (Mistral, Llama, Qwen).

The AI developer and security community must continue to actively manage these threats (as well as additional safety and security concerns) through independent testing and guardrail development throughout the lifecycle of model development and deployment in organizations. Without AI security solutions—such as multi-turn testing, threat-specific mitigation, and continuous monitoring—these models pose significant risks in production, potentially leading to data breaches or malicious manipulations. By prioritizing security-first design and layered protections, stakeholders can fortify open-weight AI systems, ensuring resilient deployments for the future.

\bibliography{main}

\end{document}